\documentclass[showpacs,twocolumn,amsmath,amssymb]{revtex4}

\usepackage{graphicx}
\usepackage{algorithm}
\usepackage{algorithmic}
\usepackage{subfigure}
\usepackage{color}

\begin{document}

\title{
  Serving by local consensus in the public service location game
}

\author{
  Yi-Fan Sun$^{1,2}$\footnote{Email: {\tt sunyifan@ruc.edu.cn}} and
  Hai-Jun Zhou$^{2,3}$\footnote{Email: {\tt zhouhj@itp.ac.cn}}
}

\affiliation{
  $^1$Center of Applied Statistics,
  School of Statistics, Renmin University of China,
  Zhong-Guan-Cun Street 59, Beijing 100872, China \\
  $^2$Key Laboratory of Theoretical Physics, Institute of
  Theoretical Physics, Chinese Academy of Sciences, Zhong-Guan-Cun East Road 55,
  Beijing 100190, China \\
  $^3$School of Physical Sciences, University of Chinese Academy of Sciences,
  Beijing 100049, China
}

\date{March 24, 2016}

\begin{abstract}
  We discuss the issue of distributed and cooperative decision-making in a
  network game of public service location. Each node of the network can choose
  to be a provider of service which is accessible to the provider itself and
  also to all the neighboring nodes. A node may also choose only to be a
  consumer, and then it has to pay a tax, and the collected tax is evenly
  distributed to all the service providers to remedy their cost. If nodes do
  not communicate with each other but make individual best-response decisions,
  the system will be trapped in an inefficient situation of high tax level.
  In this work we investigate a decentralized local-consensus selection
  mechanism, according to which nodes in need of service recommend their
  neighbors of highest local impact as candidate servers, and a node may become
  a server only if all its non-server neighbors give their assent. We
  demonstrate that this local-consensus mechanism, although only involving
  information exchange among neighboring nodes, leads to socially efficient
  solutions with tax level approaching the lowest possible value. Our results
  may help in understanding and improving collective problem-solving in various
  networked social systems and robotic systems.
\end{abstract}

\maketitle

\section{Introduction}
\label{sec:intro}

The healthy functioning of a human society depends on various public
services or facilities such as schools, hospitals, parks, garbage
disposal plants, and so on. Constructing and maintaining public service is
costly and the costs are paid by members of the whole society through tax.
On the other hand it is often the case that a service (e.g., a hospital)
located at one place will serve not only the people of this place but also
the people of neighboring places. Therefore the total cost of fulfilling the
needs of the whole society can be considerably reduced by appropriately
choosing the service provider locations. This task of
choosing the locations of public service is an important and challenging
issue faced by a modern human society, and it is also an active
research topic in the fields of algorithmic game theory \cite{Nisan-etal-2007}
and network game \cite{Bramoulle-Kranton-2007,Galeotti-etal-2010,DallAsta-Pin-Ramezanpour-2013}.

Governmental institutions may prefer to solve such public service location
problems in a top-down and centralized manner. A central planner will collect
all the needed information about the network property of the social system, and
then it will take this structural knowledge as input to a global optimization
algorithm to obtain a minimum-cost solution. But such a centralized approach
has drawbacks. Firstly it requires a central planner and requires complete
information about the networked system, and secondly the vast members of the
society are not involved in the decision process and their individual
preferences are not necessarily incorporated. This latter lack of involvement
may cause people to suspect that their interests are compromised and may then
induce strong friction and unwillingness. There are many recent large-scale
events of such types of distrust and friction. For example in 2007 some
residents in Xiamen (a major city of southeast China) protested against the
planned settlement of a PX (paraxylene) plant for fear of unexpected health and
environment effects, which eventually forced the relocation of this chemical
plant \cite{nv2007}.

A completely different approach for solving the public service location problem
is to let individual agents make choices in best response to the choices of
their neighboring agents. If an agent can access service from a neighbor in
the network, it will have no motivation to be a service provider itself, and
an agent will choose to be a service provider only if none of its network
neighbors offers service. Such a free-market approach has been investigated
in the recent literature
\cite{Bramoulle-Kranton-2007,Galeotti-etal-2010,DallAsta-Pin-Ramezanpour-2009,DallAsta-Pin-Ramezanpour-2013,Altarelli-Braunstein-DallAsta-2014},
and it was found that the resulting maximal independent-set solutions are far
from being socially efficient.

In the present work, we propose a local-consensus selection mechanism for
solving the public service location problem. This decentralized approach lies
between the above-mentioned centralized and individualized approaches. Briefly
speaking, the basic rules of this mechanism are that agents in need of service
recommend their network neighbors of highest local impact (to be precisely
defined later) as candidate service providers, and an agent may become a
service provider only if all its non-server neighbors are happy with this
arrangement. This distributed selection mechanism does not require the global
structural information of the system but only involves local-scale information
exchange. Yet very encouragingly, we find that it leads to socially efficient
solutions with tax level approaching the lowest possible value.

Our theoretical results suggest that distributed decision-making
through local consensus can be an efficient mechanism for solving the
network public service location problem. This collective mechanism may also
be useful for other network resource allocation problems
\cite{Yi-Hong-Liu-2015,Wong-Saad-2006,Yeung-Wong-2009,Cardinal-Hoefer-2006}.
In addition it may have potential applications in robotic swarm systems for
collective problem-solving \cite{Rubenstein-etal-2014} and be relevant to the
research branch of distributed algorithmic mechanism design
\cite{Nisan-Ronen-2001,Feigenbaum-Shenker-2002}.

We describe the public service location problem in the next section; then the
best-response dynamics (Sec.~\ref{sec:brd}), the centralized planning approach
(Sec.~\ref{sec:cp}), and the local-consensus mechanism  (Sec.~\ref{sec:lc}) are
discussed and their performances are compared. We conclude this work in
Sec.~\ref{sec:conc}. Some technical details are given in the two appendices.

\section{The public service location problem}
\label{sec:model}

\begin{figure}
  \begin{center}
    \includegraphics[width=0.2\textwidth]{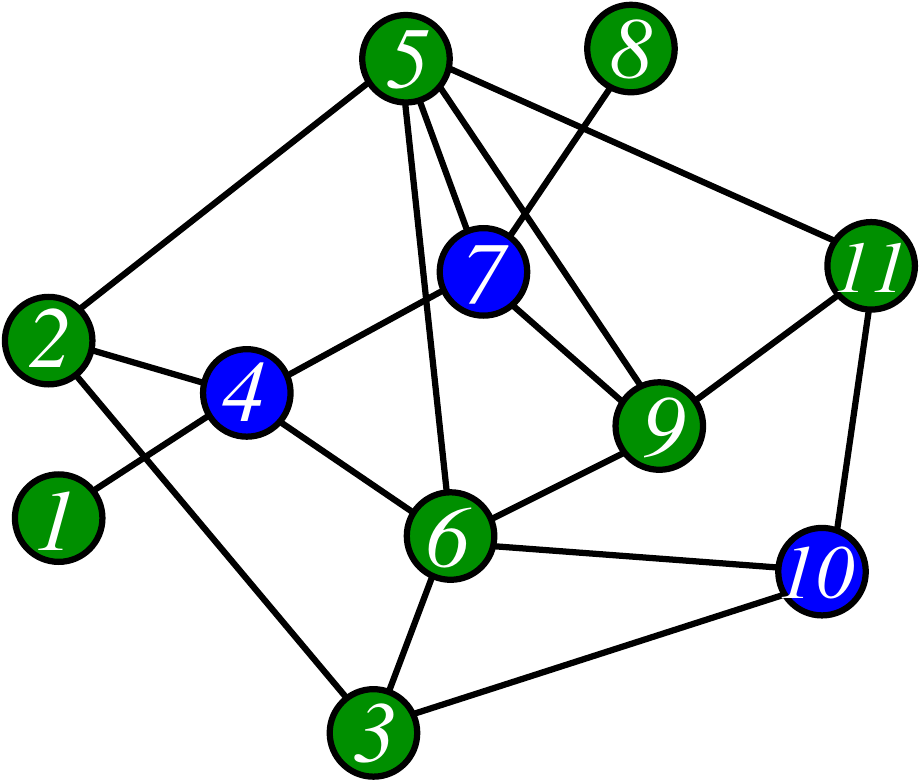}
  \end{center}
  \caption{\label{fig:pgmodel}
    An illustration of the public service location problem. There are $N=11$
    nodes (agents) and $M=16$ links in this example system. Nodes $4$, $7$, and
    $10$ are the servers and the total cost of providing service is then
    $N_1=3$. All the other eight nodes are the consumers which access service
    from the neighboring server nodes. Each consumer has to pay a tax
    $\tau =\frac{3}{11}$.
  }
\end{figure}

Let us consider a society formed by $N$ agents each of which interacting with
a set of neighboring agents. The neighborhood property is reciprocal so that if
agent $i$ is a neighbor of agent $j$ then $j$ is also a neighbor of $i$. Every
agent is dependent on certain essential public service provided by itself or by
its neighbors \cite{Bramoulle-Kranton-2007,DallAsta-Pin-Ramezanpour-2013}. We assume that the provision of
this service is costly for an agent (without loss of generality this cost is
set to be unity), but once it is provided by one agent it will be accessible to
all the neighboring agents (Fig.~\ref{fig:pgmodel}). Because of this
non-excludable nature of the public service, an agent does not need to provide
service if at least one of its neighbors is already providing it. This is
referred to as a property of strategic substitutes in the literature
\cite{Bramoulle-Kranton-2007,Galeotti-etal-2010}). If the service costs are
borne only by the service providers, naturally every agent will not volunteer
to be a provider but will wait the neighboring agents to do so, leading to
extortion and the ``tragedy of the commons''. 
The only fair solution under this cost
no-sharing rule will then be that every agent is a service provider, which is
not socially efficient as the total cost to the society is the maximum.

In this paper, therefore, we assume that the agents have reached the agreement
that free-riding is not allowed and that the total service cost is evenly
shared by all the agents in the society. The challenge faced by this society is
then a mechanism design problem: how to choose an appropriate set of agents as
service providers such that each agent is accessible to the service.

We can represent this public service location problem by a network $G$ of $N$
nodes and $M$ links. Each node of this network represents an agent and the link
$(i, j)$ between two nodes $i$ and $j$ signifies that $i$ can access the
service produced by $j$ and $j$ can access the service produced by $i$
(Fig.~\ref{fig:pgmodel}). The network structure is fixed in time, while each
node $i$ can choose to be a service provider (server, denoted by occupation
state $c_i=1$) or just be a consumer (state $c_i=0$) and it might change
between these two choices over time. A solution of this public service location
problem is then an occupation configuration
${\bf c} \equiv (c_1, c_2, \ldots, c_N)$ such that each node is either a server
(e.g., nodes $4$ and $7$ in Fig.~\ref{fig:pgmodel}) or is a consumer surrounded
by one or more servers (e.g., nodes $2$ and $6$ in Fig.~\ref{fig:pgmodel}).
The total number $N_1$ of servers in the solution ${\bf c}$ is then the total
service cost of the system, and $n_1 \equiv \frac{N_1}{N}$ is the fraction of
servers. Because of the fair-sharing rule, each consumer needs to pay a tax
$\tau = n_1$ and each server will receive a subsidy $(1-\tau)$ so as to reduce
its net cost back to $\tau$.

The question we address in this paper is:  How should the agents make decisions
about who should be the servers so that a solution ${\bf c}$ with sufficiently
low number $N_1$ of servers can be achieved? In the next section we will
demonstrate that if every agent makes choice individually and without any
cooperation, the final fraction $n_1$ of servers (and hence the tax level
$\tau$) can not be reduced below certain high level. We will then offer a
decentralized mechanism of cooperative decision-making to solve this
challenging issue efficiently.

\section{Best response dynamics}
\label{sec:brd}

Our public service location problem actually is a network game
\cite{Galeotti-etal-2010,Szabo-Fath-2007} in which each agent makes decision
under the strong self-interest of having access to the service and the weak
incentive of lowering the number of servers. A simplest decision-making
strategy is best response to the current situation of the neighborhood
\cite{Bramoulle-Kranton-2007,DallAsta-Pin-Ramezanpour-2011,Altarelli-Braunstein-DallAsta-2014}.
If an agent $i$ has one or more neighboring servers it just chooses to be a
consumer ($c_i=0$), otherwise it chooses to be a server ($c_i=1$). Starting
from an initial condition (for example, all the agents have not yet
determined), the agents update their choices non-synchronously until all are
satisfied with their last choice. After a transient period of choice changes,
this best response dynamics will converge to a solution ${\bf c}$, i.e., a Nash
Equilibrium (NE) of the game, in which all the servers are separated from each
other and every consumer has at least one neighboring server. The set of
servers of this solution ${\bf c}$ therefore is just a maximal independent set
of the network $G$ \cite{DallAsta-Pin-Ramezanpour-2009}.

\begin{figure}
  \begin{center}
    \includegraphics[angle=270,width=0.45\textwidth]{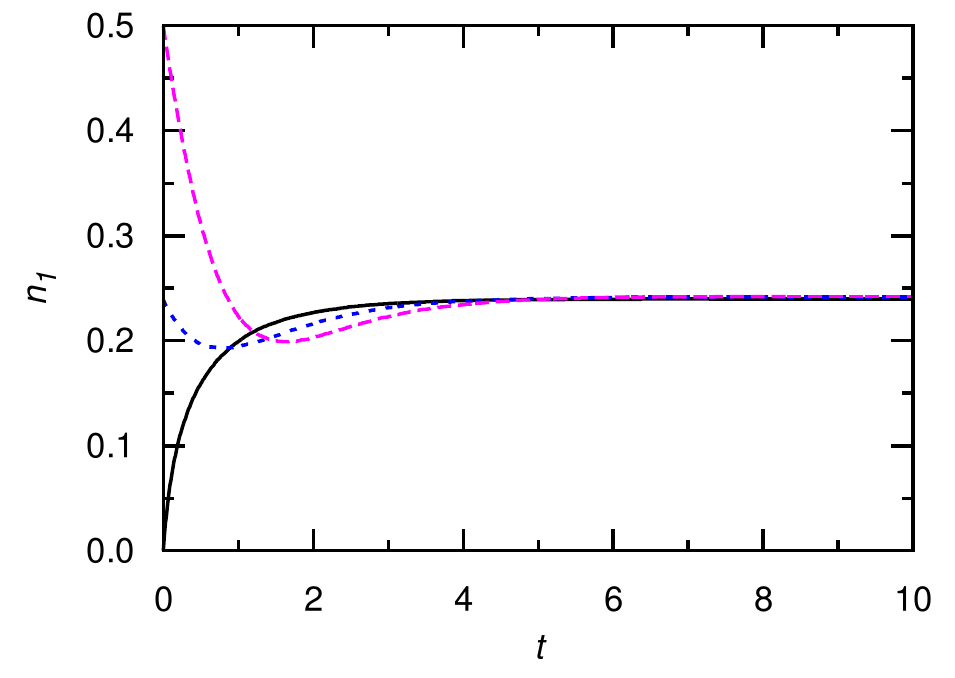}
  \end{center}
  \caption{\label{fig:IStax}
    The fraction $n_1$ of servers on a large ER random network of $N=10^5$
    nodes and mean node degree $c=10$ during the best response dynamics. At
    each time interval $\delta t=1/N$ a node $i$ is chosen uniformly at random
    from the network and its occupation state ($c_i=1$ or $c_i=0$) is then
    updated. The three curves correspond to three different initial conditions
    with fraction of servers being $0.5$, $0.24$ and $0.0$, respectively.
  }
\end{figure}

\begin{figure}
  \begin{center}
    \includegraphics[angle=270,width=0.45\textwidth]{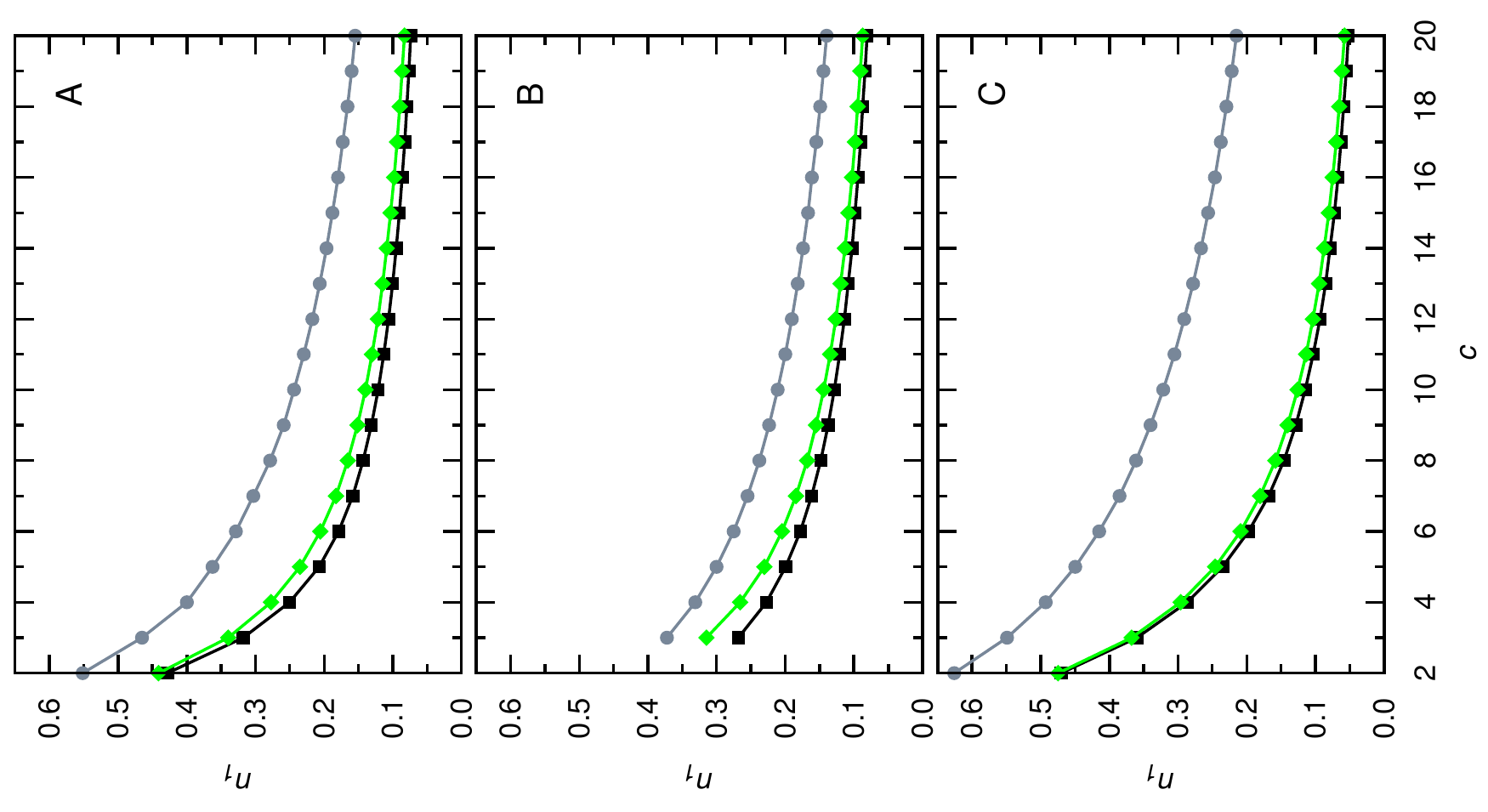}
  \end{center}
  \caption{\label{fig:ERandRRandSFjam}
    The tax level that is necessary for three different decision-making
    protocols: the best response dynamics (gray circles),
    the local-consensus dynamics (green diamonds), and the centralized
    planning (black squares). Each data point is the averaged result obtained
    on $64$ network instances with $N=10^5$ nodes and mean node degree $c$.
    We consider three types of random networks, namely ER networks (A),
    RR networks (B), and SF networks (C) with decay exponent $\gamma=3.0$
    generated through the static model \cite{Goh-Kahng-Kim-2001}.
  }
\end{figure}

Figure~\ref{fig:IStax} shows the evolution of the fraction $n_1$ of serves
on a single Erd\"os-R\'enyi (ER) random network with $N=10^5$ agents and
$M=5\times 10^5$ links (on average each agent has $c= 10$ neighbors). The $M$
links of an ER network are chosen uniformly at random from the total number
$N(N-1)/2$ of possible links. We notice that the final value of
$n_1 \approx 0.240$ reached by this best response dynamics is independent of
the initial conditions and it is in excellent agreement with the predicted
value of $n_1 = \ln(11)/10 \approx  0.2398$ by a mean field theory (see
\cite{DallAsta-Pin-Ramezanpour-2009} or Appendix~\ref{app:jamming}).
In general, the final fraction of servers reached by the best response dynamics
is $n_1 = \ln(1+c)/c$ for an ER network of mean node degree $c$
(Fig.~\ref{fig:ERandRRandSFjam}A).

The same converging behavior is observed for many other random network
instances and real-world network instances. Every node in a regular random (RR)
network has the same integer degree $K$ (so the mean degree $c = K$). For such
a network the fraction of servers converges to the final value
$n_1 = [1-(K-1)^{2/(2-K)}]/2$ (Fig.~\ref{fig:ERandRRandSFjam}B)
\cite{Wormald-1995}.  The node degrees of an exponential network obey an
exponential distribution with mean value $c$. For such a random network
we find that the fraction of servers converges to
$n_1 = [(1+3 c)^{2/3}-1]/(2 c)$ [see Eq.~(\ref{eq:jammExp})].

We also consider scale-free (SF) random networks which are better models of
real-world networked systems than ER or other homogeneous random networks
\cite{Albert-Barabasi-2002}. A SF network is very heterogeneous in the sense
that the probability $P(d)$ of a randomly chosen node to have $d$ attached
links decays with $d$ in a power-law form $P(d) \propto d^{-\gamma}$ with
exponent $\gamma > 2$. There are many highly connected nodes in such a network,
however we find that this structural property does not help to improve the
performance of the best-response mechanism (Fig.~\ref{fig:ERandRRandSFjam}C).
On the contrary, compared with ER and RR networks of the same (mean) node
degree $c$, the final server fraction needed in a SF network is even higher.
The best-response mechanism also performs poorly on real-world network
instances (Table~\ref{tab:realnetwork}).

\begin{table}
  \caption{
    Solving the public service location problem on eight real-world networks.
    For each network, we show the number of nodes $N$, the number of links $M$,
    the mean fraction $n_1$ of server nodes in the solutions obtained by three
    methods, namely the best-response (BR) dynamics, the local-consensus (LC)
    dynamics, and by the global BPD algorithm \cite{Zhao-Habibulla-Zhou-2015}.
  }
  \label{tab:realnetwork}
  \begin{tabular}{lrrlll}
    \hline
    & $N$ & $M$  & BR  & LC & BPD\\
    \hline
    Facebook \cite{McAuley-Leskovec-2014}   & $4039$    & $88234$   & $0.189$
    & $0.00248$ & $0.00248$ \\
    PowerGrid \cite{Watts-Strogatz-1998} & $4941$    & $6594$    & $0.486$
    & $0.310$   & $0.300$ \\
    CondMat   \cite{Leskovec-Kleinberg-Faloutsos-2007} & $23133$   & $93439$
    & $0.357$ & $0.158$   & $0.156$ \\
    Gnutella  \cite{Ripeanu-etal-2002} & $62586$   & $147892$
    & $0.597$ & $0.202$   & $0.201$ \\
    LocGowalla \cite{Cho-etal-2011}& $196591$  & $950327$  & $0.478$
    & $0.214$   & $0.212$ \\
    DBPL    \cite{Yang-Leskovec-2015}   & $317080$  & $1049866$ & $0.416$
    & $0.147$   & $0.147$ \\
    RoadNet-PA \cite{Leskovec-etal-2009}  & $1088092$
    & $1541898$ & $0.423$ & $0.334$   & $0.305$ \\
    YouTube \cite{Yang-Leskovec-2015}   & $1134890$ & $298624$  & $0.615$
    & $0.188$   & $0.188$\\
    \hline
  \end{tabular}
\end{table}

\section{Centralized planning}
\label{sec:cp}

If there is a central planner who has complete structural information about the
network $G$, this central planner can try to get an optimized solution
${\bf c}$ for the public service location problem by global optimization and
then appoints some agents as servers accordingly. Actually the set $\Gamma$ of
severs, with the property that every node in $G$ either belongs to $\Gamma$ or
has a neighboring node in $\Gamma$, is nothing but a dominating node set for
network $G$ \cite{Haynes-Hedetniemi-Slater-1998,Nacher-Akutsu-2016}.
Therefore an optimal service location solution corresponds to a minimum
dominating set, which has the smallest cardinality among all possible
dominating sets.

Unfortunately the minimum dominating set problem is a NP-hard
(nondeterministic polynomial hard) combinatorial optimization problem, meaning
that a guaranteed optimal solution can only be obtained by checking an
exponential number of candidate solutions. In practice one can only solve the
minimum dominating set problem approximately, and so far the best way appears
to be converting it to a spin glass model and then treating it by methods of
statistical physics \cite{Mezard-Montanari-2009}. Such a spin glass approach
can offer an estimate about the size of minimum dominating sets, and it also
offers a powerful message-passing algorithm called BPD
(belief propagation-guided decimation) for solving single network instance.
For random networks, the solutions obtained by the BPD algorithm are very close
to be minimum dominating sets \cite{Zhao-Habibulla-Zhou-2015}.
To be self-contained, some technical details of this algorithm are given in
Appendix~\ref{sec:bpd}.

Here we use the result obtained by the BPD algorithm as a good proxy of the
true optimal solution. By applying the BPD algorithm to the ER network instance
of Fig.~\ref{fig:IStax} we obtain a service location solution ${\bf c}$ with a
fraction $n_1 = 0.121$ of servers, which is much better than the solutions
($n_1 \approx 0.240$) obtained by the best-response dynamics. This result
confirms that the service location solutions obtained by the best-response
dynamics cost too much to the society. The same conclusion holds for other
random networks (Fig.~\ref{fig:ERandRRandSFjam}) and real-world network
instances (Table~\ref{tab:realnetwork}).

Although BPD or other global optimization methods can obtain socially efficient
solutions for the public service location problem, such centralized mechanism
design approaches may not be feasible in some social systems if either the
central planner is absent or the computational burden on the central planner is
unmanageable; even if they are feasible they may be unfavorable among members
of a society for fear of manipulation. Is it possible to achieve
close-to-optimal solutions for the public service location problem through
distributed planning? We give a positive answer to this question in the next
section.

\section{The local-consensus mechanism}
\label{sec:lc}

\begin{figure}
  \begin{center}
    \includegraphics[width=0.45\textwidth]{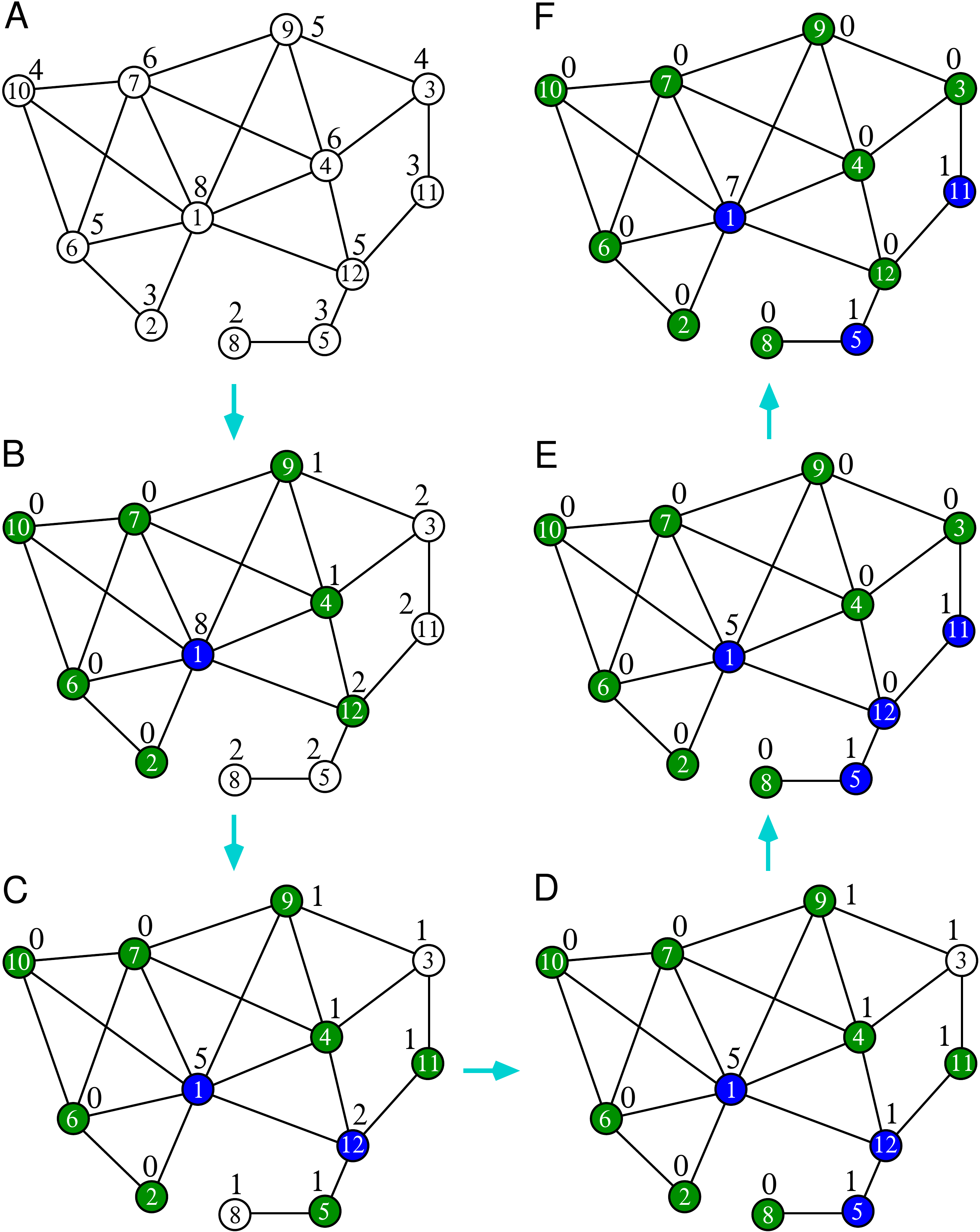}
  \end{center}
  \caption{
    \label{fig:localimpact}
    The local-consensus selection mechanism for solving the public service
    location problem distributively and cooperatively. Blue nodes are servers,
    green nodes are served consumers, and white nodes are unserved consumers.
    The non-negative integer beside a node is its impact, which might changes
    with the game process. (A) Initially there is no server node, and node $1$
    is the only candidate server who is agreed by all its neighbors. (B) After
    node $1$ changes to be a server, all its neighbors are served, and then
    $\{3, 5, 8, 11, 12\}$ becomes the set of candidates. (D)-(E) Nodes $12$,
    $5$ and $11$ are then sequentially chosen as server nodes by the
    local-consensus rule, resulting in a server arrangement with node $12$
    having zero impact. (F) Node $12$ is changed back to be a consumer,
    resulting in the final server set $\{1, 5, 11\}$, which is an optimal
    solution.
  }
\end{figure}

We now propose a local-consensus collective selection mechanism to reach a
cooperative solution for the public service location problem. Let us define the
impact $f_i$ of a node $i$ as follows: If $i$ is a server ($c_i=1$), its impact
$f_i$ is the total number of consumers which rely exclusively on $i$, i.e., the
consumers who can not access the service any longer once $i$ changes to be a
consumer; if $i$ is a consumer ($c_i=0$), its impact is the increase in the
number of served nodes if $i$ becomes a server (see Fig.~\ref{fig:localimpact}
for concrete examples). The impact of a node changes with time during the
local-consensus dynamics. We assume that every node can read the latest impact
values of all its neighbors and their latest occupation states as well.

In our local-consensus mechanism the servers for the network are assigned
sequentially until every node is being served (namely, it is either a server or
has one or more servers among its neighbors). Initially there is no server in
the network and all the nodes are unserved consumers with the impact of a node
$i$ simply being $f_i = 1 + d_i$, where the degree $d_i$ is the number of this
node's neighbors (Fig.~\ref{fig:localimpact}A). At each elementary time
interval every non-server node $i$ checks its neighborhood: if $i$ is
\emph{unserved} (having no neighboring server), then it regards a neighboring
node $j$ as suitable to be a server if and only if $j$ has the highest impact
among $i$'s neighboring nodes and $f_j$ is no less than $f_i$; if node
$i$ is \emph{served} (having at least one neighboring server) then it regards
a neighboring unserved node $k$ as suitable to be a sever if $f_k \geq f_i$.
An unserved node becomes a server candidate if it is regarded as a suitable
server by all its neighbors (e.g., node 1 in Fig.~\ref{fig:localimpact}A), 
while for a served non-server node, it becomes a server candidate if only 
all the neighboring unserved nodes recommend it as a
server (e.g., node 12 in Fig.~\ref{fig:localimpact}B). There will be one or more non-server nodes which are evaluated as
server candidates. One node (say $k$) is selected uniformly at random from these
candidates to be a server ($c_k=1$), and then all its neighbors update their
impacts and the game process repeats
(see Fig.~\ref{fig:localimpact}B-\ref{fig:localimpact}E).
After all the servers are selected through such a local-consensus mechanism,
if a server node has zero impact (e.g., node $12$ in
Fig.~\ref{fig:localimpact}E), then it is
changed back to be a consumer. This polish process is carried out
in a random sequential manner until all the remaining servers have
positive impact (Fig.~\ref{fig:localimpact}F). After the final set of
server nodes is obtained (whose relative size being $n_1$), the tax level
$\tau$ is then set to $\tau = n_1$ so that every node bears the same cost
independent of its role.

Applying this local-consensus mechanism to the ER network of
Fig.~\ref{fig:IStax}, the server fraction of obtained solutions is
$n_1 \approx 0.140$, which is a big drop as compared with the server fraction
of $n_1 \approx 0.240$ of the best-response mechanism, and it is only slightly
beyond the server fraction of $n_1 \approx 0.121$ reached by the global BPD
algorithm. This local-consensus mechanism also leads to a big drop in the
fraction of servers for other ER random network instances and RR networks and
also for scale-free random networks whose structures are very heterogeneous
(Fig.~\ref{fig:ERandRRandSFjam}).  Its performance on real-world network
instances is also very encouraging, as the solutions obtained by  
local-consensus and those obtained by the global BPD algorithm are almost
equally good in terms of server fractions $n_1$ (Table~\ref{tab:realnetwork}),
and they are much better than the solutions obtained by the best-response
dynamics.

\begin{figure}
  \begin{center}
    \includegraphics[angle=270,width=0.45\textwidth]{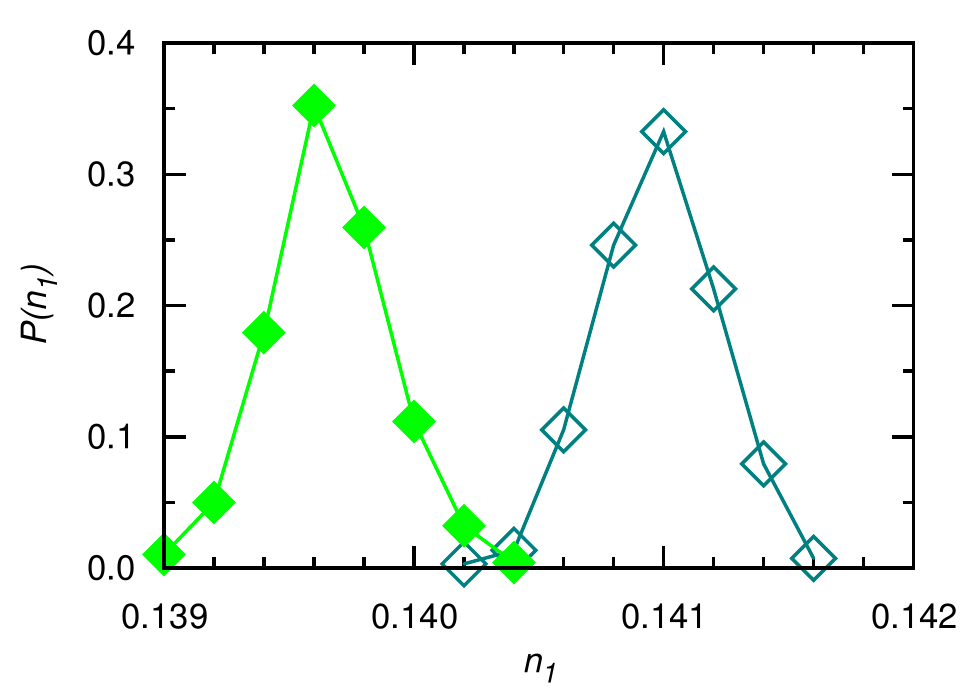}
  \end{center}
  \caption{
    \label{fig:ERimpactDiff}
    Comparing the performances of the local-consensus selection mechanism
    and the greedy highest-impact algorithm
    \cite{Haynes-Hedetniemi-Slater-1998,Molnar-etal-2013,Takaguchi-Hasegawa-Yoshida-2014} on an ER network of $N=10^5$ nodes
    and $M=5 \times 10^5$ links. Each histogram $P(n_1)$ of server fraction
    $n_1$ is obtained by sampling $960$ independent solutions.
  }
\end{figure}

Compared with the central planning approach, a nice advantage of the
local-consensus mechanism is that each node does not need to know the
structure of the whole network $G$ but only needs to know who are the
neighbors and what are their states (server, unserved or served consumer)
and current impact values. The essence of this decentralized mechanism is that
the nodes recommend their highest-impact neighbors as candidate servers.
An \emph{unserved} node will only be selected as a server if it currently has
the highest impact among its neighbors and the neighbors of its unserved
neighbors. Through this mechanism, a served consumer node may change to be a
server in response to the recommendation of all its unserved neighbors.

From the algorithmic point of view, the local-consensus mechanism is very
similar to a greedy algorithm which repeatedly selects among the whole network
a highest-impact consumer node and changes it into a server
\cite{Haynes-Hedetniemi-Slater-1998,Molnar-etal-2013,Takaguchi-Hasegawa-Yoshida-2014}.
Interestingly, we observe that the performance of the local-consensus mechanism
slightly outperforms this greedy algorithm
(Fig.~\ref{fig:ERimpactDiff}). This surprising difference can be explained by
two factors: first the local-consensus mechanism does not perform a global
ranking of nodes based on their impact values, so a node of low impact value
may become a server earlier than a node of much higher impact value; and
second, and more importantly, the local-consensus selection mechanism may
convert a served consumer $i$ to a server even if $i$ has neighbors of higher
impact values.

\section{Conclusion and discussions}
\label{sec:conc}

In this paper we considered the public service location problem as a
cooperative game among $N$ agents in a network, and presented a local-consensus
selection mechanism through which a set of high-impact agents are appointed as
service providers. We demonstrated that this decentralized selection mechanism
can reduce the societal cost of providing service to a low level that is close
to the lowest-possible value.

From the theoretical point of view, the demonstrated excellent performance of
the local-consensus mechanism is very encouraging. Our work suggests that it
is theoretically possible to efficiently solve the service location problem
by distributed decision-making. The local-consensus mechanism does not need
a central planner and it does not require the structural knowledge about the
whole network. Furthermore, every agent participates in the decision-making
process and its opinion has been incorporated in the final cooperative
solution, which may help stabilizing the solution.

For simplicity we ignored the issue of congestion in accessing service, but
this is itself an interesting factor to explore
\cite{Altarelli-Braunstein-DallAsta-2014}.
We didn't discuss the actual implementation of the local-consensus mechanism.
Instead we assumed the ideal situation that every agent is cooperative and
obeys the microscopic rules of the local-consensus mechanism. The practical
feasibility of the local-consensus mechanism is an issue to be addressed in
future empirical studies.

Collective problem-solving, division of labor, and role specialization are
common not only in human societies but also in various other social systems
such as social insects (e.g., ants and bees) and biological multi-cellular
systems \cite{Eberhart-Kennedy-1995,Kennedy-1998,Fontanari-2010} and swarms of
robots \cite{Krieger-etal-2000,Rubenstein-etal-2014}.
For robotic systems, it might be relatively easy to implement the
local-consensus decision-making mechanism to facilitate efficient division of
labor and collective problem-solving.

\section*{Acknowledgement}

This work was supported by the National Natural Science Foundation of China
(grant numbers 11121403 and 11225526), the State Key Laboratory of Theoretical
Physics (grant number Y5KF201CJ1), the Fundamental Research Funds for the
Central Universities, and the Research Funds of Renmin University of China
(grant number 14XNLF13).

\begin{appendix}

  \section{Mean field theory for the best response dynamics}
  \label{app:jamming}

  In this appendix we present a mean field theory to compute the final fraction
  $n_1$ of occupied (server) nodes under the best response dynamics. This theory
  is applicable for a generic random network with node degree distribution
  $P(d)$. It focuses on the evolution of the following quantities:
  \begin{enumerate}
  \item[]
    $\overline{N}_{u}(t)$:
    mean number of unserved nodes at time $t$;
  \item[]
    $P_{u}(d; t)$:
    probability that at time $t$ a randomly chosen unserved node has degree
    $d$;
  \item[]
    $\overline{H}_{1}(t)$:
    mean value of the sum of degrees of all the occupied nodes at time $t$;
  \item[]
    $\overline{H}_{0}(t)$:
    mean value of the sum of degrees of all the unoccupied nodes at time $t$;
  \item[]
    $\overline{H}_{u}(t)$:
    mean value of the sum of degrees of all the unserved nodes at time $t$.
  \end{enumerate}

  Initially all the nodes are unoccupied so $\overline{H}_{0}(0)= 2 M = N c$,
  where $c$ is the mean node degree. At each time interval
  $\delta t \equiv \frac{1}{N}$ a randomly chosen unserved node is occupied,
  therefore
  \begin{equation}
    \overline{H}_{0}(t+\delta t) =  \overline{H}_{0}(t) -
    \sum\limits_{d}  P_{u}(d; t) d \; .
  \end{equation}
  On the other hand, $\overline{H}_{1}(0)=0$ and
  \begin{equation}
    \overline{H}_{1}(t+\delta t) = \overline{H}_{1}(t) +
    \sum\limits_{d}  P_u(d; t) d   \; .
  \end{equation}
  The mean accumulated degree $\overline{H}_u(t)$ is expressed as
  \begin{equation}
    \label{eq:Hut}
    \overline{H}_{u}(t)= \overline{N}_u(t) \sum\limits_{d}  P_{u}(d; t) d \; .
  \end{equation}

  Initially all the nodes are unserved, $\overline{N}_u(0)=N$. At each time
  interval $\delta t$ a randomly chosen unserved node is occupied and all the
  unserved neighbors of this newly occupied node become served. Since the
  nearest neighbors of each occupied node are all unoccupied, the probability
  at time $t$ of a randomly chosen neighbor of an unserved node also being
  unserved is equal to
  $\frac{\overline{H}_u(t)}{\overline{H}_0(t)-\overline{H}_1(t)}$.
  Consequently the evolution of the number of unserved nodes is governed by
  \begin{equation}
    \overline{N}_u(t+\delta t) = \overline{N}_u(t) -1
    - \sum\limits_{d} P_u(d; t)
    \frac{\overline{H}_u(t) d}{\overline{H}_0(t) - \overline{H}_1(t)} \; .
  \end{equation}
  From this equation we can obtain the evolution equation for $P_u(d; t)$
  as
  \begin{equation}
    P_{u}(d; t+\delta t) =
    \frac{
      P_{u}(d; t) \Bigl[1-\frac{1}{\overline{N}_{u}(t)}
        - \frac{\frac{\overline{H}_{u}(t) d}{\overline{H}_{0}(t)-
            \overline{H}_{1}(t)}
        }{\overline{N}_{u}(t)} \Bigr]
    }{
      1-\frac{1}{\overline{N}_{u}(t)} -
      \frac{ \frac{\overline{H}_{u}(t)}{
          \overline{H}_{0}(t)-\overline{H}_{1}(t)}
        \sum_{d^\prime}  P_{u}(d^\prime; t) d^\prime
      }{\overline{N}_{u}(t)}
    } \; .
  \end{equation}

  Let us define several intensive quantities $h(t)$, $h_{u}(t)$, $\rho_u(t)$,
  and $c_u(t)$ as:
  \begin{subequations}
    \begin{align}
      h(t) & \equiv
      \bigl[\overline{H}_{0}(t) - \overline{H}_{1}(t)\bigr]/N \; , \\
      h_u(t) & \equiv \overline{H}_{u}(t) / N \; , \\
      \rho_u(t) & \equiv  \overline{N}_u(t) / N \; , \\
      c_u(t) & \equiv \sum_{d} P_u(d; t) d \; .
    \end{align}
  \end{subequations}
  $\rho_u(t)$ is the fraction of unserved nodes at time $t$, and $c_u(t)$ is
  the mean degree of unserved nodes at time $t$. From Eq.~(\ref{eq:Hut}) we
  know that
  \begin{equation}
    \label{eq:huexp}
    h_u(t) = \rho_u(t) c_u(t) \; .
  \end{equation}
  Furthermore, at the limit of $N\rightarrow \infty$ we have
  \begin{eqnarray}
    \frac{\partial h(t)}{\partial t}  &=& - 2 c_u(t) \; ,
    \label{eq:dhdtexp} \\
    \frac{\partial \rho_u(t)}{\partial t} &=& -1 -
    \frac{h_{u}(t)}{h(t)} c_u(t) \; ,
    \label{eq:drhouvst}
  \end{eqnarray}
  and the evolution of $P_u(d; t)$ is governed by
  \begin{equation}
    \frac{\partial P_u(d; t)}{\partial t} =
    P_u(d; t) \frac{h_u(t)}{\rho_u(t) h(t)} \bigl[ c_u(t) - d \bigr] \; .
    \label{eq:PukEvol}
  \end{equation}

  Combining Eq.~(\ref{eq:PukEvol}) and Eq.~(\ref{eq:huexp}) we can obtain an
  explicit expression for $P_u(d; t)$ as
  \begin{equation}
    \label{eq:PukEvol3}
    P_u(d; t) = \frac{P(d) \exp\bigl(-d \int_0^{t}
        \frac{c_u(t^\prime)}{h(t^\prime)} {\rm d} t^\prime
    \bigr)}{
      \sum\limits_{d^\prime}
      P(d^\prime) \exp\bigl(-d^\prime\int_0^{t} \frac{c_u(t^\prime)}{h(t^\prime)}
        {\rm d} t^\prime\bigr)} \; .
  \end{equation}
  Because of Eq.~(\ref{eq:dhdtexp}) we know that
  \begin{equation}
    \label{eq:integ1}
    \int\limits_{0}^{t} \frac{c_u(t^\prime)}{h(t^\prime)} {\rm d} t^\prime
    = -\frac{1}{2} \ln \Bigl[\frac{h(t)}{c}\Bigr] \; .
  \end{equation}
  Plugging this expression into Eq.~(\ref{eq:PukEvol3}) we finally obtain
  that
  \begin{equation}
    \label{eq:PukEvol4}
    P_u(d; t) = \frac{
      P(d) \bigl[h(t)/c \bigr]^{d/2}
    }
    {
      \sum\limits_{d^\prime} P(d^\prime)
      \bigl[h(t)/c \bigr]^{d^\prime/2}
    } \; .
  \end{equation}
  while $h(t)$ is obtained by solving the self-consistent equation
  \begin{equation}
    \label{eq:ddhvst2}
    \frac{\partial h(t)}{\partial t}
    = - 2 \frac{
      \sum\limits_{d}  P(d) d \bigl[h(t)/c \bigr]^{d/2}
    }
    {
      \sum\limits_{d^\prime} P(d^\prime)
      \bigl[h(t)/c \bigr]^{d^\prime/2}
    }
    \; .
  \end{equation}
  With $h(t)$ known, we can then obtain $c_u(t)$ from Eq.~(\ref{eq:dhdtexp})
  and then apply Eq.~(\ref{eq:drhouvst}) to obtain  $\rho_u(t)$ as the solution
  of the following differential equation
  \begin{equation}
    \frac{ \partial \rho_u(t)}{\partial t}
    = - 1 - \rho_u(t) \frac{ \bigl[c_u(t)\bigr]^2 }{ h(t)} \; .
  \end{equation}

  As time $t$ increases the fraction $\rho_u(t)$ of unserved nodes decreases
  continuously and approaches zero at certain threshold time $t^*$, i.e.,
  $\rho_u(t^*)=0$. Since a node is occupied at each time interval $\delta t$,
  the final fraction $n_1$ of occupied nodes during this best response dynamics
  is simply $n_1 = t^*$.

  In the following subsections we apply this mean field theory to several
  simple network ensembles.

  \subsection{Erd\"os-R\'enyi network}

  The degree distribution for an ER network is
  \begin{equation}
    P(d) = \frac{e^{-c} c^d}{d!} \; ,
  \end{equation}
  For this random network ensemble we have
  \begin{subequations}
    \begin{align}
      h(t) & = c (1 - t)^2 \; , \\
      c_u(t) & = c (1-t) \; , \\
      \rho_u(t) & = \frac{1+c}{c} e^{-c t} - \frac{1}{c} \; .
    \end{align}
  \end{subequations}
  Therefore the fraction of occupied nodes $n_1$ is
  \begin{equation}
    \label{eq:jammER}
    n_1 = \frac{ \ln (1+c)}{c} \; .
  \end{equation}
  Equation (\ref{eq:jammER}) was derived earlier in
  Ref.~\cite{DallAsta-Pin-Ramezanpour-2009} following the probabilistic
  approach of Ref.~\cite{Wormald-1995}.

  \subsection{Regular random network}

  In a regular random network every node has the same (integer) degree $c=K$,
  therefore
  \begin{equation}
    P(d) = \delta_d^K \; .
  \end{equation}
  For this random network ensemble we have
  \begin{subequations}
    \begin{align}
      h(t) & = K (1 - 2 t) \; , \\
      c_u(t) & = K  \; , \\
      \rho_u(t) & = \frac{K-1}{K-2} (1-2 t)^{K / 2} - \frac{1- 2 t}{K - 2} \; .
    \end{align}
  \end{subequations}
  Therefore the fraction of occupied nodes $n_1$ is
  \begin{equation}
    \label{eq:jammRR}
    n_1 =\frac{1}{2} \bigl[ 1 - (K-1)^{\frac{2}{2-K}} \bigr] \; .
  \end{equation}
  Equation (\ref{eq:jammRR}) was derived earlier in
  Ref.~\cite{Wormald-1995}.

  \subsection{Exponential random network}

  The degree distribution for an exponential random network of mean degree $c$
  is
  \begin{equation}
    P(d) = \frac{1}{1+c} \bigl(\frac{c}{1+c}\bigr)^{d} \; .
  \end{equation}
  For this random network ensemble we have
  \begin{subequations}
    \begin{align}
      h(t) & = \frac{(1+c - \sqrt{1+ 2 c t})^2}{c} \; , \\
      c_u(t) & = \frac{1+c - \sqrt{1+ 2 c t}}{\sqrt{1+ 2 c t}}  \; , \\
      \rho_u(t) & = \frac{1}{3 c} \Bigl[\frac{1+ 3 c }{\sqrt{1+2 c t}}
        -(1+ 2 c t) \Bigr] \; .
    \end{align}
  \end{subequations}
  Therefore the fraction of occupied nodes $n_1$ is
  \begin{equation}
    \label{eq:jammExp}
    n_1 =\frac{(1+ 3 c)^{\frac{2}{3}}-1}{2 c} \; .
  \end{equation}
  The correctness of Eq.~(\ref{eq:jammExp}) has been confirmed by our
  numerical simulation results.

  \section{The BPD algorithm for the service location problem}
  \label{sec:bpd}

  Selecting a minimum set of agents as service providers in a network is an
  intrinsically difficult combinatorial optimization problem. In the computer
  science literature this problem is usually referred to as the minimum
  dominating set problem. The spin glass model for the minimum dominating set
  problem and the associated replica-symmetric mean field theory have already
  been discussed in great detail \cite{Zhao-Habibulla-Zhou-2015}. Here we
  briefly review this mean field theory and the BPD message-passing algorithm.

  Given an input network $G$, the marginal probability $q_i^{c_i}$ that a node
  $i$ of this graph is in the occupation state $c_i \in \{0, 1\}$ is estimated
  by
  \begin{equation}
    \label{eq:mqi}
    q_i^{c_i}  = \frac{ e^{-x c_i} \prod\limits_{j\in \partial i}
      \sum\limits_{c_j} q_{j\rightarrow i}^{(c_j,c_i)} - \delta_{0}^{c_i}
      \prod\limits_{j\in \partial i} q_{j\rightarrow i}^{(0,0)}}
    {\sum\limits_{c_i^\prime} e^{-x c_i^\prime}
      \prod\limits_{j\in \partial i} \sum\limits_{c_j}
      q_{j\rightarrow i}^{(c_j,c_i^\prime)}
      -\prod\limits_{j\in \partial i} q_{j\rightarrow i}^{(0,0)} } \; ,
  \end{equation}
  where $x$ is a positive re-weighting parameter;
  the Kronecker symbol $\delta_{m}^{n}=1$ if
  $m=n$ and $\delta_{m}^{n}=0$ if otherwise; and $\partial i$ denotes the
  set of neighboring nodes of node $i$. For a link $(i,j)$ between two nodes
  $i$ and $j$, we denote by $q_{j\rightarrow i}^{(c_j,c_i)}$ the joint probability
  that $i$  is in occupation state $c_i$ and $j$ is in occupation state $c_j$
  when the constraint of node $i$ (that is, $i$ should
  be occupied or be surrounded by at least one occupied neighobr)
  is not considered. This `cavity' probabiity can be evaluated through the
  following belief-propagation (BP) equation:
  \begin{equation}
    q_{j\rightarrow i}^{(c_j, c_i)}  =
    \frac{ e^{-x c_j} \prod\limits_{k\in \partial j\backslash i}
      \sum\limits_{c_k} q_{k\rightarrow j}^{(c_k,c_j)} -
      \delta_{0}^{c_i+c_j} \prod\limits_{k\in \partial j\backslash i}
      q_{k\rightarrow j}^{(0,0)}}{
      \sum\limits_{c_i^\prime, c_j^\prime}
      e^{-x c_j^\prime}  \prod\limits_{k\in \partial j\backslash i}
      \sum\limits_{c_k^\prime} q_{k\rightarrow j}^{(c_k^\prime, c_j^\prime)}
      -\prod\limits_{k\in \partial j\backslash i} q_{k\rightarrow j}^{(0,0)}  } \; ,
    \label{eq:mBP}
  \end{equation}
  where $\partial j\backslash i$ denotes the subset obtained by deleting
  node $i$ from set $\partial j$.

  Equations (\ref{eq:mqi}) and (\ref{eq:mBP}) are exploited by the BPD
  algorithm to construct a near-optimal dominating set for the network $G$.
  The details of the BPD algorithm are given in \cite{Zhao-Habibulla-Zhou-2015}.
  Roughly speaking, at each round of the BPD process, first the BP equation
  (\ref{eq:mBP}) is iterated on the network a few number of times, then the
  occupation prababilities $q_i^{c_i}$ of all the unoccupied nodes $i$ are
  estimated by Eq.~(\ref{eq:mqi}), and then those nodes with the highest
  probabilities of being occupied are set to be occupied. More and more nodes
  become occupied as the BPD process continues, and it stops as soon as a
  dominating set is reached.

  The sizes of dominating sets constructed by the BPD algorithm are not
  sensitive to the re-weighting parameter $x$ \cite{Zhao-Habibulla-Zhou-2015}.
  We fix the value of $x$ to be $x=10$ in the present work.

\end{appendix}


\end{document}